\documentclass[letterpaper,english,twocolumn,prb,letterpaper,english,longbibliography,nobibnotes,nofootinbib]{revtex4}
\usepackage{mathptmx}
\usepackage[T1]{fontenc}
\usepackage[latin9]{inputenc}
\usepackage{amsthm}
\usepackage{amsmath}
\usepackage{amssymb}
\usepackage{graphicx}

\makeatletter

\pdfpageheight\paperheight
\pdfpagewidth\paperwidth

\providecommand{\tabularnewline}{\\}

\@ifundefined{textcolor}{}
{%
 \definecolor{BLACK}{gray}{0}
 \definecolor{WHITE}{gray}{1}
 \definecolor{RED}{rgb}{1,0,0}
 \definecolor{GREEN}{rgb}{0,1,0}
 \definecolor{BLUE}{rgb}{0,0,1}
 \definecolor{CYAN}{cmyk}{1,0,0,0}
 \definecolor{MAGENTA}{cmyk}{0,1,0,0}
 \definecolor{YELLOW}{cmyk}{0,0,1,0}
}
\numberwithin{equation}{section}
\numberwithin{figure}{section}

\makeatother

\usepackage{babel}
\begin{document}

\title{Statistical Common Author Networks (SCAN)}

\author{F.G. Serpa}

\affiliation{Booz Allen Hamilton Inc., 3811 N. Fairfax Dr., Ste. 600, Arlington,
Virginia 22203}

\author{Adam M. Graves}

\affiliation{Booz Allen Hamilton Inc., 3811 N. Fairfax Dr., Ste. 600, Arlington,
Virginia 22203}

\author{Artjay Javier}

\email{javier_artjay@bah.com }

\affiliation{Booz Allen Hamilton Inc., 3811 N. Fairfax Dr., Ste. 600, Arlington,
Virginia 22203}
\begin{abstract}
A new method for visualizing the relatedness of scientific areas is
developed that is based on measuring the overlap of researchers between
areas. It is found that closely related areas have a high propensity
to share a larger number of common authors. A methodology for comparing
areas of vastly different sizes and to handle name homonymy is constructed,
allowing for the robust deployment of this method on real data sets.
A statistical analysis of the probability distributions of the common
author overlap that accounts for noise is carried out along with the
production of network maps with weighted links proportional to the
overlap strength. This is demonstrated on two case studies, \emph{complexity
science} and \emph{neutrino physics}, where the level of relatedness
of areas within each area is expected to vary greatly. It is found
that the results returned by this method closely match the intuitive
expectation that the broad, multidisciplinary area of \emph{complexity
science} possesses areas that are weakly related to each other while
the much narrower area of \emph{neutrino physics} shows very strongly
related areas.
\end{abstract}
\maketitle

\section{Introduction}

Understanding the growth and evolution of academic research areas\cite{Borner2004_KnowledgeMapIntro,Gulbahce2010_MapSci}
is important to assessing the health and influence of scientific areas
and can provide potentially important predictive capability in assessing
technologies that may emerge from fundamental and applied research.
A consequence of the large and growing number of highly specialized
research areas is that identifying the productive intersection of
these\cite{Porter2009_Interdisciplinary} can no longer be done manually.
However, the ready availability of computing power, the large frequency
of published work and the relatively high data integrity of bibliographic
databases provides the elements necessary for automated screening
and visualization of these interdisciplinary areas.

The visualization of research areas is an active area in bibliometric
studies\cite{Gulbahce2010_MapSci,chen2001_VisKnowledge}, largely
using clustering of individual units to describe the relatedness of
research areas. The primary metric conferring this relatedness has
historically been the citation frequency\cite{ioannidis2006_CitationFreq},
with the individual unit of measure being an instance of one publication
citing another. Using publications as nodes, a very complicated unweighted
directed network could be formed relating publications together. Since
this is visually confusing, the practice of re-assigning these nodes
as either authors or journals\cite{nerur2005_CitationNet} is preferable,
resulting in a weighted network map where the clustering observed
in these networks broadly reflects the topical areas of study, creating
a variation of what is traditionally referred to as a knowledge map.
Intuitively, these methods work well at understanding the relatedness
of topical areas because authors tend to cite research in the area
of their study more frequently and journals tend to publish work that
caters to a specific, topically focused scientific community. Of value
in these visualizations are the areas of study that lie between topical
clusters that represent interdisciplinary research, which can often
give rise to emerging scientific areas.

While the methods described above use citation as the fundamental
unit of measure, we offer an alternative approach by showing how counting
the occurrences of the same author working in multiple areas can provide
the necessary linking to relate these multiple areas to each other.
Intuitively, this approach is motivated by the observation that scientists
working in one area of study will work in related areas of study more
frequently than in unrelated areas, and so we expect a stronger connection
between closely related areas. The approach we carried out produces
an undirected, weighted network map that differs from the practice
described above in the following ways: (1) the nodes themselves are
the topical areas of study, (2) the weight of the link connecting
one node to the next is proportional to the number of authors shared
between those topical areas (adjusted for area size), and (3) the
clustering observed will define a major topical area composed of closely
related topics. In general, the network constructed in this way can
be thought of as a bipartite graph where one set of nodes corresponds
to authors and the other set to the areas that the authors participate
in. In principle, the structure of this network is amenable to analysis
with many clustering or community detection algorithms. One of the
values of using common authors over citations is that the links observed
are much stronger since they require authors to develop deep expertise
in these areas in order to publish successfully in them, as opposed
to a simple understanding of the work executed, which is the minimum
requirement to cite another's work in the case of citation patterns.
In this paper, we develop the procedures for establishing this link,
in particular correcting for name homonymy in a statistical way.

This approach is also useful for examining interdisciplarity, which
we define here to be a feature arising from the participation of two
or more vastly different areas of study in terms of expertise, knowledge
or training. Our concept of relatedness is measured by the overlap
in participation of researchers in two different areas. Large overlaps
in participation between two or more areas indicates strong relatedness.
Low overlaps in participation between two or more areas indicates
poor relatedness. In a very simple example, if two areas belong to
separate clusters with little or no relatedness to each other, but
both areas have strong relatedness with a third area, that is indicative
that the third area is interdisciplinary since it draws on the participation
of two, unrelated areas.

For our case studies to demonstrate this methodology we intentionally
selected two different fields: the \emph{complexity sciences} and
\emph{neutrino physics}. While the latter is a traditional, narrow
field of study that is deeply rooted in physics, the former is a multidisciplinary
field that intersects with many other scientific areas, drawing upon
the talents of many different types of scientists. For this reason
we intuitively anticipate a stronger degree of relatedness in \emph{neutrino
physics} than \emph{complexity science}, and show that the method
we developed confirms that intuition. Last, we point out that while
a significant number of methods in bibliometrics focus on relationships
between authors and papers (i.e. co-authorship\cite{Newman2001_CoAuthorNet}
or citation patterns) that elucidate the structure and pattern \emph{within}
these areas, this approach focuses on the relationships \emph{between}
these areas.

\section{Methodology}

The publications used to generate the common author graphs were drawn
from the Institute for Engineering and Technology's Inspec publication
database as accessed through the Thompson Reuter's ISI Web of Knowledge
v5.5 index. Once the Inspec database was selected through the Web
of Knowledge search interface the Boolean keyword or series of keywords
best representing the field under investigation were entered into
the Inspec search field. For clarity, the term sub-field will be used
for these specific searches, where it is understood that the keyword
search was structured in such a way as to extract a scientific community
that is engaged in studying a sub-field (i.e. social cybernetics)
that happens to also be part of a larger field of study (i.e. complexity
science). The generic term ``area'' will be used when the context
could conceivably pertain to both field and sub-field. This is a subject
matter expert managed process. In general, keywords that are most
closely associated with a field of study were selected such that it
would conservatively capture papers within the field of interest.
There is some variance between keywords and spot-checking the articles
by a subject matter expert within the area was used to validate that
each keyword pull consisted of only relevant articles. However, the
method described here is not limited to keyword searches and can be
applied to classification indices, journals, university research output,
or any arbitrarily chosen group of articles.

The search was performed over the years 1969-2012, the longest time
span available in the database, however the vast majority of searches
returned results with shorter durations. Each keyword search typically
returned $10{}^{1}$ to $10{}^{5}$ publications. A custom Python
script was written and used to pre-process the database by sub-fields
to extract a list of authors, where the last name and initials were
stored, and repetitions were removed. This produced a list of unique
authors for every keyword search. These lists were then compared with
each other to determine the number of authors the lists had in common.
A symmetric matrix of pairwise comparisons was generated in this way
using fast search algorithms in Python. Typical computing times were
on the order of a few minutes for the generation of individual topics
lists, while the overlap between topics required on the order of several
hundred searches over the sub-fields and took approximately half an
hour, using server-class hardware.

\section{Discussion}

As a first approximation to quantifying the link between any two fields
of study one can postulate the number of authors common to both fields.
Unfortunately this naive approach suffers from two deficiencies that
precludes its use as a measure of overlap: area size dependence and
noise. Intuitively, it can be reasoned that the number of common authors
depends in some way on (1) the number of authors in each topic, which
varies by several orders of magnitude based on area size, and (2)
the probability of false positives that arise from matching two authors
that are different people with the same last name and initials. These
occurrences, though rare, cannot be eliminated easily and are globally
present and mostly uniform. For these reasons they will be referred
to as noise arising from name homonymy, which is a persistent problem
in bibliometrics \cite{TorviK2009_Homynymy,lagoze2011_Homynymy}.
Below, we develop a treatment for both of these effects.

\subsection{An Equation to Handle Multiple Fields of Different Sizes}

First, we develop a treatment to deal with the large variation of
area sizes that will affect the number of common authors in the pairwise
matching. In what follows we try to derive an expression for the number
of matches as a fuction of list sizes and how they relate to the probability
of finding name matches. We have not found a simple exact derivation
of a formula relating these quantities but procedurally we offer a
formula and motivate it using some simplifying assumptions and show
how the said formula is justified for our purposes by comparing its
results to a Monte Carlo simulation.

Let us consider a pool of names and from it extract two lists of names,
$N$ and $M$, containing $n$ and $m$ elements respectively and
with no loss of generality assume that $n\leq m$, and that the names
be unique within the lists, but not necessarily between each other.
Let us start by comparing one element of $N$ to one of the elements
in the list of $M$ and further assume that there exists a probability
$p$ for an element of $N$ to be matched to an element of $M$. There
are two outcomes: the element either matches that entry in the list
with probability $p$, or it does not (with probability $1-p$). Since
there are $m$ elements in $M$, the probability of finding no matches
between the first element in $N$ to the \emph{entire} list of $M$
will be $(1-p)^{m}$. However, we are not interested in the case of
no matches, but in the case of matches, that can now be approximated
by: $1-(1-p)^{m}$, as the probability that a single element in $N$
will match an element in list $M$ (strictly speaking the last expression
represents also the case of multiple matches but we assume the chance
to be small and actually precluded by the assumption that each list
has no internal matches). Now we proceed to develop an expression
for comparing the entirety of both lists to each other. As a first
approximation we can multiply the probability of the single element
matching case by the number of elements, $n$, to produce the expression
in Equation 1, where $E(k)$ is the expected number of matches between
the lists of size $n$ and $m$ (this value $\overline{k}$ will be
later approximated by the number of matches obtained from real data)

\begin{equation}
E(k)=n\left(1-(1-p)^{m}\right)\label{eq:k}
\end{equation}

For our purposes, the unknown variable is $p$. In order to make use
of Equation \ref{eq:k}, we rewrite the variables from their expected
values to their measured values. Thus, $\overline{k}$ will be an
estimate of $E(k)$, and $\overline{p}$ will be an estimate of $p$.
Solving for $\overline{p}$ produces the functional form of the equation
we will use.

\begin{equation}
\overline{p}=1-\left(1-\frac{\overline{k}}{n}\right)^{\frac{1}{m}}\label{eq:p}
\end{equation}

Equation \ref{eq:k} is just an approximation and worth noting how
it may fail and in what regime. First as we compare lists every time
there is a match the second list should be reduced by $1$ and the
probabilities should be adjusted accordingly. This could be accounted
for by trying to perform an exact calculation or perhaps by intuitevely
postulating an effective $m*$ that is somehow smaller than $m$.
In practice, we expect that since $p$ and the numbers of matches
are small, the formula will still be a valid approximation. Notice
also that we do not expect the formula to be symmetric with respect
to $n$ and$m$ since we have assumed that $n\leqslant m$. 

In order to validate our use of these approximations, we carried out
a Monte Carlo (MC) simulation of the exact solution over the range
of $n$ and $m$ within the lists used in this study by generating
matches between list for different values of $p$, computing an expected
number of matches $\bar{k}$, and trying to recover the initial value
of $p$ by using Equation \ref{eq:p}. The MC simulation takes two
integers $n$ and $m$ and with probability $p$ generates matches.
This procedure introduces the nuances we do not treat in our derivation,
like the fact that once a match is found then the largest list is
reduced. The MC simulation also calculates only single matches as
opposed to accounting for multiple matches as in the assumptions above.
The result is that for sizes within the ranges used, there was less
than 5\% error between the Monte Carlo result and the analytical expression
on Equation \ref{eq:p}, supporting our use of the latter as a valid
approximation. Briefly, an example calculation looking at two fields
consisting of 1000 authors ($n$) and 10,000 authors ($m$) that happens
to have 100 matches $\overline{k}$ between them allows us to use
Equation \ref{eq:p} to calculate the probability $(p)$ to be $1-\left(1-\frac{100}{1000}\right)^{\frac{1}{10000}}=1.05\times10^{-5}$.
As a check, we compare this to what the MC simulation would predict
as the number of matches given the same $N$, $M$, $p$ and obtain
$98.8$ matches while Equation \ref{eq:p} would predict $99.6$ matches,
representing an error of less than $1\%$.

\subsection{An Approximation to Noise Arising from Name Homonymy}

Now that an expression for the matching probability has been developed,
it can be used as a measure of the strength of the link between various
areas of study, which describes the overall probability that authors
in one area will also publish in the paired area. While that is the
focus of this study, it is first important to characterize the amount
of noise arising from name homonymy. The statistics arising from the
matching probabilities as calculated from the number of matched authors
(the pairwise comparison matrix described in the methodology) can
be used to determine this noise factor. To do this, we choose pairs
of fields in which we intuitively expect to find no true overlap of
common authors, implying that the overlap found is due solely to name
homonymy. Specifically, we apply Equation 2 to the pairwise comparison
of 25 fields within \emph{neutrino physics} to 25 fields within \emph{complexity
science}. This produces a matrix of values of matching probability
that describe the occurrence of name homonymy. We plot the histogram
of these values in Figure 1, Top. The histogram shows a very broad,
skewed distribution of probabilities with a second delta-like distribution
centered at zero. The median value of this distribution is $p=1.62\times10^{-6}$,
which is represenatative of name homonymy since the fields compared
in this way have very little relationship to each other. The broader
distribution is related to the name homonymy error, whereas the delta-like
peak at zero is a result of matches from lists with very small numbers
of authors where zero is a very likely outcome. In order to validate
this against ground truth, the list of authors identified in this
way was randomly spot-checked by slecting 20 authors at random and
using the open source search engine Google along with the affiliations
listed in their papers to find the specific individuals. It was verified
that all of the common authors found in this way corresponded to two
or more distinct individuals, thus lending support to our assertion
that this is a reasonable method to estimate the level of name homonymy.
A larger and broader sampling will help establish the estimated error,
but based on the result that the 20 selected had no homonymy errors,
it is expected errors generated in this way range from 0-9\% within
a 95\% confidence interval.

\begin{figure}
\includegraphics[height=0.5\paperheight]{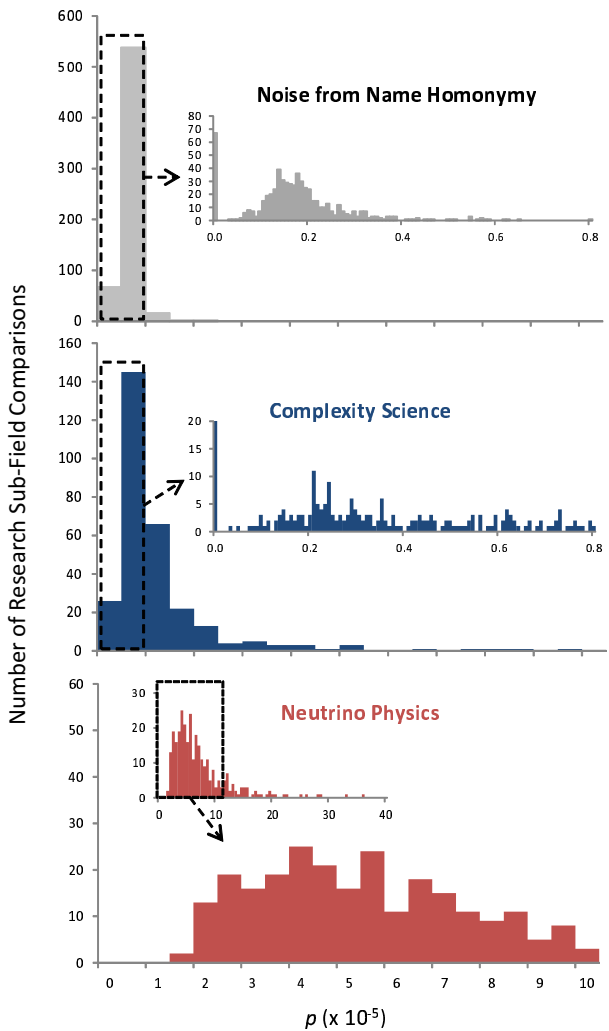} \caption{Histograms of scientific fields surveyed plotted as a function of
the matching probability between fields of study$(p)$ as calculated
in Equation 2, using bins of the same size for comparison. Top: Probability
values between areas in\emph{ neutrino physics} and \emph{complexity
science} is representative of name homonymy error. Middle: Probability
values for \emph{complexity science}. Bottom: Probability values for
\emph{neutrino physics}. Note that insets in Top and Middle show higher
resolution (more bins over a smaller range) while the inset of the
Bottom shows wider range (more bins but also at a much wider range).}
\end{figure}

\subsection{Case Study: Neutrino Physics vs. Complexity Science}

Similar statistical analyses were then carried out on the areas of
\emph{neutrino physics} and \emph{complexity science}, comparing fields
within each area inclusively. In Figure 1 Middle, a histogram plotting
the matching probability values of the pairwise comparison matrix
of \emph{complexity science} is shown. It can be seen that while there
is a large number of matching probability values that correspond to
the peak of the name homonymy noise, there are a significant number
of matches that far exceed these values. The median value of this
distribution is $p=4.13\times10^{-6}$ and is representative of the
amount of name homonymy plus participation of complexity authors in
multiple sub-fields. Still, its similarity in the peak of the distribution
to noise suggests that this is a very weakly related area of study
where there are very few common areas between fields. For example,
if we divide the medians in this way to simulate a signal-to-noise
ratio (S/N), we obtain $4.13\times10^{-6}/1.62\times10^{-6}=2.55$
which is generally considered to be a very weak signal. This matches
well with our intuition and knowledge of \emph{complexity science}
which tends to be strongly interdisciplinary, drawing scientists working
in diverse areas such as sociology, biology, computer science and
economics. Complexity scientists do not share common skillsets, training,
or equipment and there remains debate on the defining elements and
boundaries of their field.

In Figure 1 Bottom, a similar treatment is carried out for the field
of \emph{neutrino physics}. Here we find strong overlap between authors
as exemplified by a shift in the distribution toward much higher matching
probability values. The median value of this distribution is $p=5.38\times10^{-6}$
and is representative of the amount of name homonymy plus participation
of neutrino authors in multiple sub-fields. This is more than an order
of magnitude larger than the complexity science median. This indicates
that the field of \emph{neutrino physics} is very strongly related,
with a large number of scientists in one sub-field publishing in many
others. Using a similar signal-to-noise argument as complexity science,
it is found to be S/N=33, which is generally considered to be a relatively
strong signal. This also matches our intuition since we know that
this area of study is very deeply rooted in physics, requiring very
expensive specialized instruments and a much smaller, less diverse
physics-oriented community. Physicists studying neutrinos have a very
similar skillset and training and in fact not only use similar but
sometimes the same equipment.

Using the medians of the distributions as a measure of the differences
in relatedness of the two sub-fields, we calculateed the ratio (subtractively
corrected for name homonymy noise) to be 22.4 times more likely for
an author to publish in multiple sub-fields if they are in neutrino
science than in complexity science. A complete summary of these statistical
calculations appears in Table \ref{tab:stats}.

\begin{table*}[t]
\caption{Statistics of the distribution of matching probabilities $(p)$ complexity
science, neutrino physics and the name homonymy are shown comparitively.
Signal-to-noise ratios $(S/N)$ are given for each field (CS or NP)
with respect to using name homonymy as the noise, $N$. Additionally,
treating the noise as a background, the actual signal can also be
obtained subtractively $(S-N)$. It is also useful to compare how
much of a difference a highly relatedess field is to a low relatedness
field by looking at the ratio.}

\begin{tabular}{|c|cc|c|c|}
\cline{2-5} 
\multicolumn{1}{c|}{} & \multicolumn{1}{c|}{Median $(\times10^{-6})$} & Mean $(\times10^{-6})$ & Median$(S/N)$ & Median$(S-N)$ \tabularnewline
\hline 
Name Homonymy (N) & $1.62$ & $1.80$ & \multicolumn{1}{c}{--} & --\tabularnewline
\cline{1-1} \cline{4-5} 
Complexity Science (CS) & $4.13$ & $11.0$0 & $2.55$ & $2.33$\tabularnewline
\cline{1-1} 
Neutrino Physics (NP) & $53.8$ & $69.4$ & $33.2$ & $52.18$\tabularnewline
\hline 
Ratio (NP/CS) & -- & \multicolumn{1}{c}{--} & -- & $22.39$\tabularnewline
\hline 
\end{tabular}\label{tab:stats}
\end{table*}

Now we use the statistics gathered to define the link strength $(l)$
to be the matching probability $(p)$ between fields \emph{within}
\emph{complexity science} and \emph{neutrino physics} minus the mean
of the matching probability$(p_{0})$ of the name homonymy \emph{between}
\emph{complexity science} and \emph{neutrino physics},

\begin{equation}
l=p-\overline{p_{0}}
\end{equation}

A plot of the fields of \emph{complexity science} and \emph{neutrino
physics} are shown in Figure 2, where a higher link strength is represented
by the thicker line weights for the lines connecting each node. 

\begin{figure}
\includegraphics[height=0.5\paperheight]{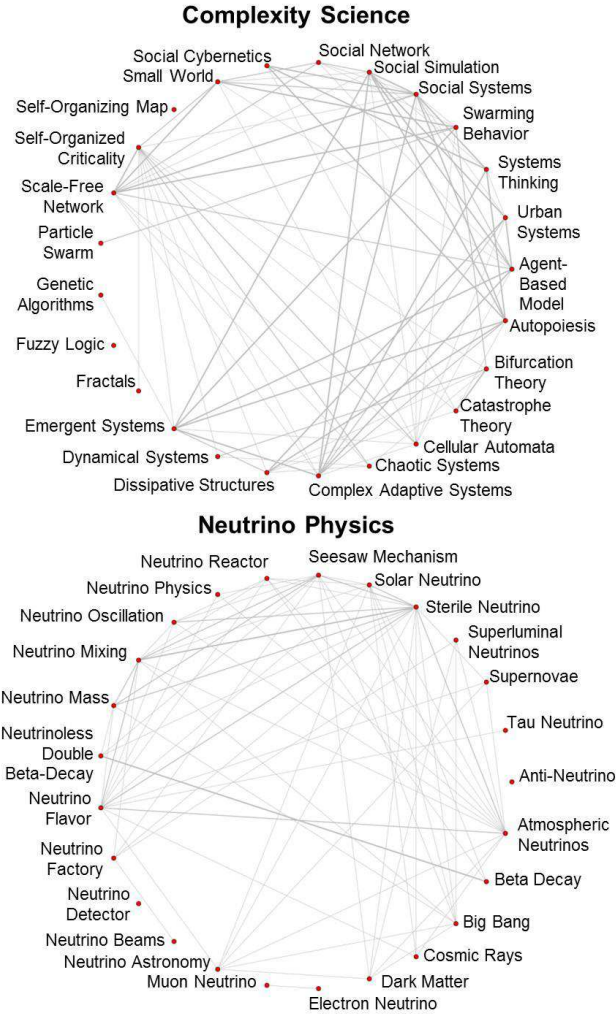}

\caption{Network structure of the fields of \emph{complexity science} (top)
and \emph{neutrino physics} (bottom) showing the relatedness of sub-fields
of study (nodes) as determined from the number authors that are common
to each sub-field of study (adjusted for sub-field size and corrected
for name homonymy noise). The thickness of the lines represents the
link weight and is proportional to the matching probability between
sub-fields. Note that for clarity, the link weights are consistent
within the top and bottom figures but not relative to each other;
if done this way, then the lines in the top graph will be too faint
to see.}
\end{figure}

We observe additional intuitive verification when looking at the relative
link weighting. For example in \emph{complexity science}, very thick
weighted lines connect the social related areas: \emph{social network,
social simulation, social systems, social cybernetics}. Additionally,
areas where there is little connection also bears out our expectations.
For example, the only sub-field connected to \emph{particle swarm}
is \emph{swarming behavior}, as expected. Sub-fields which are subsets
of each other also possess strongly weighted links as expected. For
example in \emph{neutrino physics}, there is a very strong link between
\emph{beta decay} and \emph{neutrinoless double beta decay}, as papers
(and therefore authors) of the former sub-field also contain papers
from the latter since the keyword of the former is included in the
latter.

\subsection{Other applications of SCAN}

For this case study, we have shown that this method produces results
that we intuitively expect, in order to validate the underlying assumptions
concerning the area participation of authors. In general, however,
this method can be applied to any arbitrary grouping scheme. As this
method uses the overlap of authors within groupings, it allows the
user to test the validity of using that particular grouping scheme.
For example, a popular mapping method for scientific knowledge is
to use citation patterns between journals, where the clusters indicate
areas of study. A similar map of science can be generated using the
SCAN method, where each node is a journal and each edge between any
two journals is proportional to the number of authors that published
in both journals. Comparing and contrasting these two methods can
allow for settling questions which cannot be addressed by either method
alone.

It is often of value to simplify arbitrarily chosen areas in maps
of knowledge by using the smallest number of distinct elements, since
it is possible that some of these elements are synonymous or extensions
of other elements. The SCAN method can be applied to these problems
as a filter to weed out areas with high relatedness to other areas
(or possibly induce these areas to be merged together). Thus a knowledge
map that is progressively simplified in this way will have its mean
link strength gradually approach a minimum value, as the areas with
large amounts of relatedness are removed from the map. 

SCAN can also be applied to emerging research areas, where it is important
to understand the underlying communities that these emerging areas
arise from and the authors in this emerging research area can be compared
with their participation in other areas. This concept can be taken
further by incorporating more graph-centric concepts like clustering.
An emerging area that is simply an extension of other areas will display
strong mutual participation from those underlying areas, while a truly
distinct emerging area will have low participation from the underlying
areas.

We have described the development and execution of a novel method
in visualizing scientific areas by looking at common authors, which
is valuable in the study of emerging interdisciplinary areas. The
links of this network are a function of how easily the methods and
training in one area can contribute to work in a related area. Compared
to other knowledge mapping methods that look at patterns in citations
and collaborations, this method is much more selective as common authorship
can only occur when an author has a depth of expertise to allow him
or her to publish original work in multiple areas. A comparison of
mapping differences between these measures and applications of community
detection algorithms is planned for future work. Last, since our development
of an approach to estimate the expected noise arising from name homonymy
and the further statistical treatment of establishing link weightings
resulted in an approximation, future work is planned to explore this
more accurately. It is possible to use the same level of name homonymy
as calculated in this study as a correction factor for other studies
since it should be measuring a global phenomenon, but the method outlined
here provides way forward to estimate this more accurately by including
more disparate areas if that is desired.

\section{Acknowledgments}

The authors would like to thank Todd Hylton (Brain Corporation, former
DARPA PM) for his support, as well as their colleagues at Booz Allen
Hamilton, Inc. for their support and useful discussions: Jennifer
Klamo, Zigurts Majumdar, David Guarrera, J. Tyler Whitehouse, Marie
Sandrock and Allan Steinhardt.

\section{References}

\bibliographystyle{unsrt}
\bibliography{SCAN}

\end{document}